\documentclass[aps,prd,twocolumn,showpacs,nofootinbib]{revtex4}

\usepackage{amsfonts}
\usepackage{mathrsfs}
\usepackage{graphicx}
\usepackage{amsmath}
\usepackage{amssymb}

\begin{document}

\title{$J/\psi +\chi_{cJ}$ Production at the $B$ Factories under the Principle of Maximum Conformality}

\author{Sheng-Quan Wang}

\author{Xing-Gang Wu}
\email[email:]{wuxg@cqu.edu.cn}

\author{Xu-Chang Zheng}

\author{Jian-Ming Shen}

\author{Qiong-Lian Zhang}

\address{ Department of Physics, Chongqing University, Chongqing 401331, P.R. China}

\date{\today}

\begin{abstract}

Under the conventional scale setting, the renormalization scale uncertainty usually constitutes a systematic error for a fixed-order perturbative QCD estimation. The recently suggested principle of maximum conformality (PMC) provides a principle to eliminate such scale ambiguity in a step-by-step way. Using the PMC, all non-conformal terms in perturbative expansion series are summed into the running coupling, and one obtains a unique, scale-fixed, scheme-independent prediction at any finite order. In the paper, we make a detailed PMC analysis for both the polarized and the unpolarized cross sections for the double charmonium production process, $e^{+} + e^{-} \to J/ \psi (\psi') + \chi_{cJ}$ with $(J=0,1,2)$. The running behavior for the coupling constant, governed by the PMC scales, are determined exactly for the specific processes. We compare our predictions with the measurements at the $B$ factories, BABAR and Belle, and the theoretical estimations obtained in the literature. Because the non-conformal terms are different for various polarized and unpolarized cross sections, the PMC scales of these cross sections are different in principle. It is found that all the PMC scales are almost independent of the initial choice of renormalization scale. Thus, the large renormalization scale uncertainty usually adopted in the literature up to $\sim40\%$ at the NLO level, obtained from the conventional scale setting, for both the polarized and the unpolarized cross sections are greatly suppressed. It is found that the charmonium production is dominated by $J=0$ channel. After PMC scale setting, we obtain $\sigma(J/\psi+\chi_{c0}) = 12.25^{+3.70}_{-3.13}$ fb and $\sigma(\psi'+\chi_{c0}) =5.23^{+1.56}_{-1.32}$ fb, where the squared average errors are caused by bound state parameters as $m_c$, $|R_{J/\psi}(0)|$ and $|R'_{\chi_{cJ}}(0)|$, which are non-perturbative error sources in different to the QCD scale setting problem. In comparison to the experimental data, a more accurate theoretical estimation shall be helpful for a precise testing of QCD and for determining whether there is new physics beyond the Standard Model. \\

\noindent {\bf keywords}: perturbative QCD, heavy quarkonium production, QCD scale setting

\end{abstract}


\maketitle

\section{introduction}

Perturbative quantum chromodynamics (pQCD) calculations are essential to describe processes with large momentum transfer. The double charmonium production processes, $e^{+}+e^{-}\rightarrow J/ \psi \;(\psi') + \chi_{cJ}$ with $(J=0,1,2)$, play an important role for understanding the production and hadronization properties of the heavy quarkonium. In addition to being a demonstration of the non-relativistic QCD (NRQCD) factorization \cite{nrqcd}, such kind of charmonium production also provides a good opportunity to learn pQCD color-singlet mechanism~\cite{npb172425}.

In the literature, these processes have been studied within the pQCD factorization up to next-to-leading order (NLO), cf. Refs.\cite{br:2005lo,liu:2003lo1,liu:2003lo2,Zh:2008nlo,W:2011nlo,Dong:2011nlo}. Their polarized cross sections up to NLO can be written as
\begin{eqnarray}
\sigma^{J}_{\lambda_1,\lambda_2} &=& \sigma^J_{\lambda_1,\lambda_2}\left|_{\rm LO}\right. + \sigma^J_{\lambda_1,\lambda_2}\left|_{\rm NLO}\right. \nonumber\\
&=& A^J_{\lambda_1,\lambda_2}\; \alpha^{2}_{s}(\mu_R)\; \left\{1+{\alpha_s(\mu_R) \over \pi} B^J_{\lambda_1,\lambda_2}(\mu_R) \right\} , \label{pol:cross:section}
\end{eqnarray}
where $\lambda_1$ and $\lambda_2$ stand for the helicities of the outgoing $J/\psi$ and $\chi_{cJ}$. The tree-level coefficients
\begin{eqnarray}
A^J_{\lambda_1,\lambda_2} &=& {32\pi e_c^2 \alpha^2 C_F^2 \over 3 s^2 m_c^6} \left({|{\bf P}|\over \sqrt{s}}\right) |R_{J/\psi}(0)|^2 |R'_{\chi_{cJ}}(0)|^2 \nonumber\\
&& \times\; r^{(1+|\lambda_1+\lambda_2|)} \left| c^J_{\lambda_1,\lambda_2}(r) \right|^2 ,
\end{eqnarray}
where the charm quark electric charge $e_c={2/3}$ and the color factor $C_F$= $4/3$. The dimensionless parameter $r={4m^{2}_{c} / s}$, and $|{\bf P}|=\lambda^{1\over 2}(s,M^2_{J/\psi},M^2_{\chi_{cJ}}) / (2\sqrt{s})$, where
\begin{displaymath}
\lambda(x,y,z)=x^2+y^2+z^2-2xy-2yz-2zx .
\end{displaymath}
The parameters $|R_{J/\psi}(0)|$ and $|R'_{\chi_{cJ}}(0)|$ are the radial wavefunction at the origin and the first derivative of the radial wavefunction at the origin for $J/\psi$ and $\chi_{cJ}$, respectively. The LO coefficients $c^J_{\lambda_1,\lambda_2}$ and the NLO coefficients $B^J_{\lambda_1,\lambda_2}(\mu_R)$ are put in the Appendix. The parameter $\mu_R$ stands for the renormalization scale.

For calculating all the polarized and the unpolarized cross sections of the processes $e^{+}+e^{-}\rightarrow J/ \psi \;(\psi') + \chi_{cJ}$, one needs to introduce the renormalization scale $\mu_R$. In the literature, it is usually taken as the typical momentum flow of the process ($Q$), e.g. $Q=2m_c$ or $\sqrt{s}/2$ or $\sqrt{s}$, respectively. However which one can result in the right theoretical estimation is not clear. As for conventional scale setting, the renormalization scale is fixed once it has been set to be $Q$; for convenience, we call it the initial renormalization scale $\mu_R^{\rm init}$. That is, under conventional scale setting, one always sets $\mu_R\equiv\mu_R^{\rm init}=Q$. It has been observed that under such scale setting, one usually obtains sizable renormalization scale dependent estimation \cite{br:2005lo,liu:2003lo1,liu:2003lo2,Zh:2008nlo,W:2011nlo,Dong:2011nlo}, which could be up to $40\%-50\%$ for taking $\mu_R$ to be the above mentioned three typical scales, and up to $10\%-20\%$ by varying $\mu_R$ within the region of $[Q/2,2Q]$ for a specific $Q$. Thus, the renormalization scale uncertainty usually provides a large systematic error under the conventional scale setting. Especially, the arbitrary choice of $[Q/2,2Q]$ could be misleading in certain cases. For example, Ref.\cite{Q6} argues that after including the first and second order corrections to the deep inelastic sum rules which are due to heavy flavor contributions, the renormalization scale $\mu_{R}$ should be taken as $\mu_R \sim 6.5 \,m$, if taking the typical scale $Q$ to be the corresponding heavy quark mass $m$.

Recently, it has been pointed out that the principle of maximum conformality (PMC) provides a possible solution for eliminating such renormalization scale ambiguity~\cite{pmc1,pmc2,pmc3,pmc4,pmc5,pmc6,pmc7}. PMC provides the underlying principle for the BLM mechanism~\cite{blm}, a recent review on PMC can be found in Ref.\cite{pmc8}.

The main idea of PMC lies in that one can first finish the renormalization procedure for any pQCD process by using an initial value for the renormalization scale ($\mu_R^{\rm init}$), and then set the effective or optimal PMC scale for the process. The PMC scale is generally different from $\mu_R^{\rm init}$, which is formed by absorbing all non-conformal terms that rightly governs the running behavior of the coupling via the renormalization group equation into the coupling constant~\cite{pmc2}. In different to conventional scale setting, one can choose any arbitrary value to be $\mu^{\rm init}_R$, but the optimal PMC scale and the resulting finite-order PMC prediction are both to high accuracy independent of such arbitrariness, consistent with the renormalization group invariance. The PMC satisfies all self-consistency conditions for setting the renormalization scale~\cite{pmc5}. After proper procedures, all non-conformal $\{\beta_i\}$-terms in the perturbative expansion are summed into the running coupling so that the remaining terms in the perturbative series are identical to that of a conformal theory. The QCD predictions from PMC are then independent of renormalization scheme, inversely, such scheme independence can be adopted to derive commensurate scale relations among different observables~\cite{csr}. Moreover, after PMC scale setting, the divergent ``renormalon" series $(n!\;\beta_i^{n}\alpha_s^n)$ does not appear in the conformal series and the convergence of the pQCD series can be greatly improved in principle. For example, such merits have been shown by a next-to-next-to-leading order (NNLO) PMC analysis of the top-quark pair production at the hadronic colliders~\cite{pmc3,pmc4}. As the main purpose of the present paper, we shall show that even at the NLO level, the pQCD prediction after PMC scale setting can also have such good features, thus our understanding of $e^+ + e^- \rightarrow J/\psi + \chi_{cJ}$ can be greatly improved.

The remaining parts of the paper are organized as follows. In Sec.II, we give the polarized and unpolarized cross sections for the double charmonium production under PMC scale setting. Numerical results and discussions are presented in Sec.III. The final section is reserved for a summary.

\section{polarized and unpolarized cross sections}

To set the PMC scales for high-energy processes, one needs to decompose the perturbative coefficients at each $\alpha_s$ order into $\{\beta_i\}$-dependent and independent parts respectively. For the double charmonium production at the NLO level, we need to decompose the NLO coefficient $B^J_{\lambda_1,\lambda_2}(\mu^{\rm init}_R)$ into $\beta_0$-dependent part (non-conformal part) and $\beta_0$-independent part (conformal part), i.e.
\begin{equation}
B^J_{\lambda_1,\lambda_2}(\mu^{\rm init}_R) = B^{J(\beta)}_{\lambda_1,\lambda_2}(\mu^{\rm init}_R) \beta_0 + B^{J(con)}_{\lambda_1,\lambda_2}(\mu^{\rm init}_R) ,
\end{equation}
where the non-conformal $B^{J(\beta)}_{\lambda_1,\lambda_2}(\mu^{\rm init}_R)$ and the conformal $B^{J(con)}_{\lambda_1,\lambda_2}(\mu^{\rm init}_R)$ can be found in the Appendix. Then, following the standard procedure of PMC, the polarized cross section (\ref{pol:cross:section}) can be rewritten as,
\begin{widetext}
\begin{eqnarray}
\sigma^{J}_{\lambda_1,\lambda_2} &=& A^J_{\lambda_1,\lambda_2} \;\alpha^{2}_{s} \left(\mu^{\rm PMC}_{R,(\lambda_1,\lambda_2)}\right)\times \left \{1+{\alpha_s \left(\mu^{\rm PMC}_{R,(\lambda_1,\lambda_2)}\right) \over \pi} B^{J(con)}_{\lambda_1,\lambda_2}\left(\mu^{\rm init}_R\right) \right\}
\end{eqnarray}
\end{widetext}
with the PMC scale
\begin{equation}\label{pmcscale}
\mu^{\rm PMC}_{R,(\lambda_1,\lambda_2)}=\mu^{\rm init}_R \exp{\left( {-B^{J(\beta)}_{\lambda_1,\lambda_2}\left(\mu^{\rm init}_R\right)} \right)} .
\end{equation}
It shows that the PMC scales are determined by the non-conformal terms which govern the dominant behavior of the running coupling constant. At present, because the non-conformal terms $B^{J(\beta)}_{\lambda_1,\lambda_2}(\mu^{\rm init}_R)$ are different for $J=0,1,2$, the PMC scales, and hence the theoretical estimations, for these cross sections are also different. This is one of the important features of PMC scale setting.

Similarly, the NLO coefficients $B^J_{t}(\mu^{\rm init}_R)$ of the unpolarized/total cross sections can also be decomposed into the non-conformal part $B^{J(\beta)}_{t}(\mu^{\rm init}_R)$ and the conformal part $B^{J(con)}_{t}(\mu^{\rm init}_R)$. After PMC scale setting, the total cross section
\begin{eqnarray}
\sigma^J_{t} &=& \sigma^J_{t}(\mu^{\rm init}_R) \left|_{\rm LO}\right. \times \nonumber\\
&& \left\{1+{\alpha_s(\mu^{\rm init}_R)\over\pi}\left( B^{J(\beta)}_{t}(\mu^{\rm init}_R) \beta_0 + B^{J(con)}_{t}(\mu^{\rm init}_R) \right)\right\} \nonumber
\end{eqnarray}
reduces to
\begin{equation}\label{pmctotal}
\sigma^J_{t}=\sigma^J_{t}(\mu^{\rm PMC}_{R;t,J})\left|_{\rm LO}\right. \left \{1+{\alpha_s(\mu^{\rm PMC}_{R;t,J}) \over\pi} B^{J(con)}_{t}\left(\mu^{\rm init}_R\right) \right\}
\end{equation}
with the PMC scale,
\begin{equation}
\mu^{\rm PMC}_{R;t,J}=\mu^{\rm init}_R \exp{\left(-B^{J(\beta)}_{t}(\mu^{\rm init}_R)\right)} .
\end{equation}
It is noted that the initial scale dependent logarithmic terms in $B^{J(con)}_{\lambda_1,\lambda_2} \left(\mu^{\rm init}_R\right)$ and $B^{J(con)}_{t}\left(\mu^{\rm init}_R\right)$ should be absorbed into the coupling constant simultaneously with the non-conformal terms via the renormalization group equation of the coupling constant, so after applying PMC scale setting, $B^{J(con)}_{\lambda_1,\lambda_2}$ and $B^{J(con)}_{t}$ are still at the initial scale $\mu^{\rm init}_R$. As for the functions of the total cross sections (\ref{pmctotal}), we have
\begin{widetext}
\begin{eqnarray}
\sigma^0_{t}\left|_{\rm LO}\right. &=& \sigma^0_{0,0} \left|_{\rm LO}\right. + 2 \sigma^0_{1,0}\left|_{\rm LO}\right., \\
B^{0(con)}_{t} &=& \frac{1} {\sigma^0_{t} \left|_{\rm LO}\right.} \left(B^{0(con)}_{0,0} \sigma^0_{0,0}\left|_{\rm LO}\right. +2B^{0(con)}_{1,0} \sigma^0_{1,0}\left|_{\rm LO}\right. \right), \\
B^{0(\beta)}_{t} &=& \frac{1}{\sigma^0_{t}\left|_{\rm LO}\right.} \left(B^{0(\beta)}_{0,0} \sigma^0_{0,0}\left|_{\rm LO}\right. +2B^{0(\beta)}_{1,0} \sigma^0_{1,0}\left|_{\rm LO}\right. \right)
\end{eqnarray}
for $J=0$;
\begin{eqnarray}
\sigma^1_{t}\left|_{\rm LO}\right. &=&2 \left(\sigma^1_{1,0}\left|_{\rm LO}\right. +\sigma^1_{0,1}\left|_{\rm LO}\right. + \sigma^1_{1,1}\left|_{\rm LO}\right. \right) ,
\\
B^{1(con)}_{t}&=&\frac{1} {\sigma^1_{t}\left|_{\rm LO}\right.} \left(2B^{1(con)}_{1,0} \sigma^1_{1,0}\left|_{\rm LO}\right. + 2B^{1(con)}_{0,1} \sigma^1_{0,1}\left|_{\rm LO}\right. + 2B^{1(con)}_{1,1}\sigma^1_{1,1}\left|_{\rm LO}\right. \right) , \\
B^{1(\beta)}_{t} &=& \frac{1}{\sigma^1_{t}\left|_{\rm LO}\right.} \left(2B^{1(\beta)}_{1,0} \sigma^1_{1,0}\left|_{\rm LO}\right. + 2B^{1(\beta)}_{0,1} \sigma^1_{0,1}\left|_{\rm LO}\right. +2B^{1(\beta)}_{1,1} \sigma^1_{1,1}\left|_{\rm LO}\right. \right)
\end{eqnarray}
for $J=1$;
\begin{eqnarray}
\sigma^2_{t}\left|_{\rm LO}\right. &=& \sigma^2_{0,0}\left|_{\rm LO}\right. +2\sigma^2_{1,0}\left|_{\rm LO}\right. +2\sigma^2_{0,1}\left|_{\rm LO}\right. +2\sigma^2_{1,1}\left|_{\rm LO}\right. +2\sigma^2_{1,2}\left|_{\rm LO}\right. , \\
B^{2(con)}_{t} &=& \frac{1}{\sigma^2_{t}\left|_{\rm LO}\right.} \left( B^{2(con)}_{0,0}\sigma^2_{0,0}\left|_{\rm LO}\right. +2B^{2(con)}_{1,0} \sigma^2_{1,0}\left|_{\rm LO}\right. + 2B^{2(con)}_{0,1}\sigma^2_{0,1}\left|_{\rm LO}\right. \right. \nonumber\\
&& \qquad \left. + 2B^{2(con)}_{1,1} \sigma^2_{1,1}\left|_{\rm LO}\right. +2B^{2(con)}_{1,2}
\sigma^2_{1,2}\left|_{\rm LO}\right. \right) , \\
B^{2(\beta)}_{t} &=& \frac{1} {\sigma^2_{t}\left|_{\rm LO}\right.} \left( B^{2(\beta)}_{0,0} \sigma^2_{0,0}\left|_{\rm LO}\right. +2B^{2(\beta)}_{1,0} \sigma^2_{1,0}\left|_{\rm LO}\right. +2B^{2(\beta)}_{0,1}\sigma^2_{0,1}\left|_{\rm LO}\right. + 2 B^{2(\beta)}_{1,1}\sigma^2_{1,1}\left|_{\rm LO}\right. \right. \nonumber\\
&&\qquad \left. +2B^{2(\beta)}_{1,2}\sigma^2_{1,2}\left|_{\rm LO}\right. \right)
\end{eqnarray}
for $J=2$, respectively. Here, for convenience, we have omitted the initial scale $\left(\mu^{\rm init}_R\right)$ dependence to the coefficients $B^{J(con)}_{t}$ and the PMC scale $\left(\mu^{\rm PMC}_{R;t,J}\right)$ dependence in the LO cross sections.
\end{widetext}

The summed up non-conformal functions $B^{J(\beta)}_{t}(\mu^{\rm init}_R)$ are usually different from their polarized terms $B^{J(\beta)}_{\lambda_1,\lambda_2}(\mu^{\rm init}_R)$, then the PMC scales for the total cross sections are generally different from the PMC scales for the polarized cross sections. Furthermore, because the non-conformal terms $B^{J(\beta)}_{t}(\mu^{\rm init}_R)$ are different for $J=0,1,2$, the PMC scales of these total cross sections are also different. Thus, for the considered processes, we have to introduce several PMC scales for the polarized and the unpolarized cross sections. The physical picture for the introducing of different PMC scales is clear, all PMC scales are determined by those known $\{\beta_i\}$-terms that rightly determine the running behavior of the strong coupling constant in each case.

It is noted that these PMC scales only formally depend on the choice of the initial renormalization scale, but as will be shown later, all these PMC scales are almost independent of the initial choice of renormalization scale, and then the renormalization scale dependence can be greatly suppressed (or even eliminated in comparison to the errors caused by other uncertainty sources as $m_c$, $|R_{J/\psi}(0)|$, $|R'_{\chi_{cJ}}(0)|$ and etc.). As a consistent cross-check of PMC estimation, it is helpful to show whether after PMC scale setting, by summing up all the polarized cross section, one can still get the same result as that of the directly calculated unpolarized cross section. This, inversely, can also be used as a criteria of whether the complex higher-order calculations are right or not.

At the NLO level, it is noted that the coefficients proportional to $n_f$ or $\beta_0$ are the same, so one can practically obtain $\mu^{\rm PMC}_{R} = \mu^{\rm BLM}_{R}$. But there are advantages for using PMC other than BLM:
\begin{itemize}
\item The PMC provides the underling principle for BLM, which clearly shows how the $n_f$-terms are eliminated. The PMC can be extended to any perturbative order in a self-consistent way.
\item That is, to set the optimized scale, it is not simply to set the coefficient before $n_f$ to zero but to eliminate $n_f$ in a combined form as $\beta_0$ or other $\{\beta_i\}$-functions that rightly governs the running behavior of the coupling constant. Otherwise, one may not get the correct conformal term for estimating the physical observable, especially in some specific cases when the conformal and non-conformal terms have no connection~\cite{pmc5}.
\item By using PMC scale setting, one can conveniently show how the pQCD convergence can be improved in principle and how the renormalization scale dependence can be greatly suppressed or eliminated even at the NLO level.
\end{itemize}

\section{numerical results and discussions}

\subsection{Input parameters}

To do numerical calculation, we set the collision energy $\sqrt{s}=10.6$ GeV, the charm quark mass $m_c=1.5$ GeV \footnote{We adopt the so called 1S-mass for $m_c$ \cite{1Smass1,1Smass2}, i.e. half of the $J/\Psi$-mass for charm quark, which is consistent with the bound-state parameters under the potential-model calculations~\cite{wforigin1,wforigin2}. } and $\alpha(\sqrt{s})=1/130.9$ \cite{fine:130fine}. Using the two-loop $\alpha_s$ running with $\alpha(M_{Z})=0.1184$~\cite{pdg}, we obtain $\Lambda^{(n_f=3)}_{\rm QCD}=0.386$ GeV, $\Lambda^{(n_f=4)}_{\rm QCD}=0.332$ GeV, and $\Lambda^{(n_f=5)}_{\rm QCD}=0.231$ GeV. Based on the experimental values for the leptonic width of $J/\psi$ and $\psi(2S)$ and the width of $\chi_{c2}$ to two photons, we can inversely determine the radial wavefunction $|R_{ns}(0)|$ at the origin and the first derivative of the radial wavefunction at the origin $|R^{'}_{np}(0)|$ through the NLO formulas~\cite{pdg}
\begin{equation}
\Gamma_{\psi(ns)\rightarrow e^+e^-} = {4\alpha^2\over 9m^2_c}\left(1-{16\alpha_s(2m_c) \over 3\pi}\right)|R_{ns}(0)|^2 \label{wave1}
\end{equation}
and
\begin{equation}
\Gamma_{\chi(np)\rightarrow \gamma\gamma} = {64\alpha^{2}\over 45m^{4}_{c}}\left(1-{16\alpha_{s}(2m_c) \over 3\pi}\right)|R^{'}_{np}(0)|^{2}. \label{wave2}
\end{equation}
To be consistent with the present NLO analysis of quarkonium pair production, we need a NLO determination of $|R_{ns}(0)|$ and $|R^{'}_{np}(0)|$. To set the PMC scales for those decay widths, one needs to finish the NNLO calculation for these processes, which however are not available at present due to its complexity. For simplicity, we set its scale as the one usually suggested in the literature, i.e. $\mu_R=2m_c$, and only provide a rough estimation on different scale choices. For experimental values of these decay widths, we adopt those from the Particle Data Group~\cite{pdg}: $\Gamma_{J/\psi\rightarrow e^+e^-} =(5.55\pm0.16)$ keV, $\Gamma_{\psi'\rightarrow e^+e^-} =(2.37\pm0.04)$ keV and $\Gamma_{\chi_{c_2}\rightarrow \gamma\gamma}=(0.514\pm0.062)$ keV. Then, as a combined error being the squared average of the experimental errors on the decay widths and the theoretical errors caused by varying the scale within the conventional region of $[m_c,4m_c]$, we obtain:
\begin{eqnarray}
|R_{J/\psi}(0)|^{2} &=& \left(0.855^{+0.044}_{-0.051}\right) \;{\rm GeV}^{3} , \label{wfval1}\\
|R_{\psi^{'}}(0)|^{2} &=& \left(0.365^{-0.017}_{-0.020}\right) \;{\rm GeV}^{3} , \label{wfval2} \\
|R^{'}_{\chi_{cJ}}(0)|^{2} &=& \left(0.056^{+0.007}_{-0.007}\right) \;{\rm GeV}^{5} , \label{wfval3}
\end{eqnarray}
where we have neglected the spin-effects in both the same level $S$-wave and $P$-wave states, and treat all $1P$-wave states have the same $|R^{'}_{1p}(0)|=|R^{'}_{\chi_{cJ}}(0)|$. At present, in order to provide a relatively reliable estimation of the scale error, following the idea of PMC, we first transform  Eqs.(\ref{wave1},\ref{wave2}) into general renormalization scale dependent forms by including the dominant log-terms, which have been eliminated by directly setting the renormalization scale to be the typical momentum transfer $(2m_c)$~\cite{wangsq}. That is, before analyzing the scale error, Eqs.(\ref{wave1},\ref{wave2}) are rewritten by using conventional running behavior of the coupling constant as
\begin{eqnarray}
&&\Gamma_{\psi(ns)\rightarrow e^+e^-} = {4\alpha^2\over 9m^2_c}|R_{ns}(0)|^2 \times\nonumber\\
&&\quad \left[1-{16\alpha_s(\mu_R) \over 3\pi}\left(1+\beta_{0}\ln\left(\frac{\mu_R^2}{\mu_{R,0}^2}\right) \frac{\alpha_s(\mu_R)}{4\pi}\right)\right] ,\\
&&\Gamma_{\chi(np)\rightarrow \gamma\gamma} = {64\alpha^{2}\over 45m^{4}_{c}}|R^{'}_{np}(0)|^{2}\times \nonumber\\
&&\quad \left[1-{16\alpha_{s}(\mu_R) \over 3\pi}\left(1+\beta_{0}\ln\left(\frac{\mu_R^2}{\mu_{R,0}^2} \right) \frac{\alpha_s(\mu_R)}{4\pi}\right)\right] ,
\end{eqnarray}
where $\beta_0=11-2n_f/3$ and $\mu_{R,0}=2m_c$. Some subtle points need to be mentioned: I) This scale error analysis follows the idea of PMC but is an approximation because of lacking strict NNLO $\{\beta_i\}$-terms to determine its PMC scales; II) Only part of the two-loop correction that partly determines the running behavior of coupling constant has been considered, since it only involves the log-term $\ln\left({\mu_R^2}/{\mu_{R,0}^2} \right)$ dependent $\beta$-functions. Even though such log term is at the one-order higher, it is necessary and can largely compensate the scale changes at the NLO level, otherwise, one will obtain abnormally large scale errors. III) Since the choice of typical momentum flow is not unique, different choice of it may lead to extra scale uncertainties. IV) In some sense, the present idea of including the dominant scale running effect determined by the renormalization group equation into the scale error analysis is consistent with the idea of the principle of minimum sensitivity (PMS)~\cite{pms}, in which the renormalization scale is so set as to minimize the sensitivity of the estimation to the scale variation. That is, we observe that the scale choice of $\mu_R=2m_c$ corresponds to a steady point for those decay widths. This also agrees with the observation of Ref.\cite{pmc8}, i.e. even though the PMS does not satisfy the reflexivity, symmetry and transitivity properties of the renormalization group as PMC does, the PMS is consistent with PMC at least at the NLO level in which its NLO coefficients are also free from $\beta$-terms.

These values for the wavefunction parameters are consistent with those of Ref.\cite{W:2011nlo} and those derived from the potential models such as Refs.~\cite{wforigin1,wforigin2,wforigin3,wforigin4} within reasonable errors. At present, the wavefunction parameters $|R_{ns}(0)|$ and $|R^{'}_{np}(0)|$ appear in the amplitude as a linear factor, so the uncertainties for the charmonium production channels can be figured out straightforwardly, thus, throughout the paper if not specially stated, we shall fix their values to be their central values \footnote{A better determination of those parameters, such as the consideration of quark mass effect, relativistic effect, higher order effect and etc., shall be helpful for deriving an accurate estimation. Such an analysis is out of the range of present paper.}.

\subsection{Numerical results and discussions}

\begin{table}[h]
\begin{tabular}{|c||c|c|c||c|c|c|}
\hline
 & \multicolumn{3}{c||}{Conventional scale setting}  &  \multicolumn{3}{c|}{PMC scale setting}  \\
\hline
~~$\mu^{\rm init}_{R}$~~ & ~~$2\;{\rm m_{c}}$~~ & ~~${\sqrt{s}/ 2}$~~ & ~~$\sqrt{s}$~~ & ~$2\;{\rm m_{c}}$~ & ~${\sqrt{s}/ 2}$~ & ~$\sqrt{s}$~ \\
\hline\hline
$\sigma^{0}_{t}$ (fb) & 9.31 & 6.87 & 5.26 & 12.25 & 12.25 & 12.25 \\
\hline
$\sigma^{1}_{t}$ (fb) & 1.02 & 0.85 & 0.71 & 1.00 & 1.00 & 1.00 \\
\hline
$\sigma^{2}_{t}$ (fb) & 1.54 & 1.27 & 1.04 & 1.58 & 1.58 &1.58 \\
\hline
\end{tabular}
\caption{Initial renormalization scale dependence for the total cross sections of $e^{+}+e^{-}\rightarrow J/\psi+ \chi_{cJ}$, where three typical initial scales are adopted. It shows that the total cross sections strongly depend on the choice of the (initial) scale under conventional scale setting; while the total cross sections after PMC scale setting are almost independent of $\mu^{\rm init}_{R}$. } \label{scaleun}
\end{table}

\begin{figure}[t]
\includegraphics[width=0.45\textwidth]{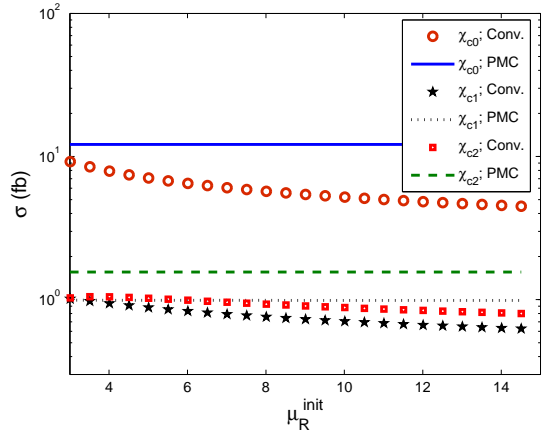}
\caption{Total cross sections versus the initial scale $\mu^{\rm init}_{R}$ for $e^{+}+e^{-}\rightarrow J/ \psi + \chi_{cJ}$ with $(J=0,1,2)$. After PMC scale setting, the total cross sections are almost flat versus $\mu^{\rm init}_{R}$. Here the short notation Conv. stands for the total cross section derived under conventional scale setting. } \label{Plot:sqrst:a}
\end{figure}

We present the total cross sections before and after PMC scale setting in Table \ref{scaleun}, where three typical momentum flows are adopted as the initial scale $\mu^{\rm init}_{R}$. Under conventional scale setting, $\mu_R\equiv \mu^{\rm init}_{R}$, the total cross sections are quite sensitive to the choice of renormalization scale. For example, by varying $\mu^{\rm init}_R$ from $2m_c$ to $\sqrt{s}/2$, total cross sections for the cases of $\chi_{c0}$, $\chi_{c1}$ and $\chi_{c2}$ production are changed by about $26\%$, $16\%$ and $8\%$ respectively. This observation agrees with previous estimations obtained in the literature~\cite{br:2005lo,liu:2003lo1,liu:2003lo2,Zh:2008nlo,W:2011nlo,Dong:2011nlo}. Then, the scale uncertainty always constitutes an important error source for the conventional estimation. We do not know which scale could be the answer unless by comparing with data, which however greatly suppresses the predictive power of pQCD. On the other hand, after PMC scale setting, we observe that the total cross sections for $\chi_{c0}$, $\chi_{c1}$ and $\chi_{c2}$ production remain almost unchanged by varying $\mu^{\rm init}_{R}$ to be disparate ones as $2{\rm m_{c}}$, ${\sqrt{s}/2}$ and $\sqrt{s}$. More explicitly, we present the total cross section versus $\mu^{\rm init}_{R}$ in Fig.(\ref{Plot:sqrst:a}). It shows that, the total cross sections are almost flat versus the initial renormalization scale, then the renormalization scale ambiguity is eliminated even at the NLO level. After PMC scale setting, we have resummed the non-conformal $\{\beta_0\}$-terms known at the NLO level into the coupling constant, generally, the LO cross sections will be significantly increased, while the NLO corrections are suppressed to a certain degree. Then, the perturbative convergence can be improved, in principle. The large $\beta_0$-approximation in this sense is consistent with our present treatment, in which all $\beta_0$-terms have been summed up, e.g. for the quarkonium electromagnetic annihilation decays~\cite{chen}. It is noted that the PMC scale setting can be conveniently extended up to any perturbative order, and any other type of $\beta$-terms can also be summed up in a consistent way.

\begin{table*}[htb]
\begin{tabular}{|c|c|c|c|}
\hline
 ~~  ~~  &~~~\textsc{Belle}~~~ & ~~~\textsc{BaBar}~~~  & ~~~our prediction~~~ \\
 ~~ ~~   & $\sigma\times\mathcal{B}_{>2 (0)}$~\cite{be:2004data} & $\sigma\times\mathcal{B}_{>2}$~\cite{ba:2005data} & ~~ ~~ \\
\hline \hline
$\sigma(J/\psi+\chi_{c0})$ & $6.4\pm1.7\pm1.0$ & $10.3\pm 2.5^{+1.4}_{-1.8}$ & $12.25^{+2.94+2.24}_{-2.26-2.17}$ \\
$\sigma(J/\psi+\chi_{c1})$ & - & -  & $1.00^{+0.37+0.18}_{-0.27-0.18}$ \\
$\sigma(J/\psi+\chi_{c2})$ & - & -  & $1.58^{+0.40+0.28}_{-0.33-0.28}$ \\
$\sigma(J/\psi+\chi_{c1}) + \sigma(J/\psi+\chi_{c2})$ & $ < 5.3$ at 90\% CL & -  &  $2.58^{+0.77+0.46}_{-0.60-0.46}$ \\
\hline
$\sigma(\psi'+\chi_{c0})$ & $12.5\pm3.8\pm3.1$ & - &  $5.23^{+1.25+0.93}_{-0.97-0.90}$  \\
$\sigma(\psi' + \chi_{c1})$  & - & - & $0.43^{+0.15+0.08}_{-0.12-0.07}$ \\
$\sigma(\psi'+\chi_{c2})$    & - & - & $0.67^{+0.18+0.12}_{-0.14-0.12}$ \\
$\sigma(\psi'+\chi_{c1}) + \sigma(\psi'+\chi_{c2})$ & $< 8.6$ at 90\% CL & - &  $1.10^{+0.33+0.20}_{-0.26-0.19}$ \\
\hline
\end{tabular}
\caption{Comparison of our theoretical predictions for the total cross sections (in unit: fb) of $J/\psi(\psi')+\chi_{cJ}$ ($J=0, 1, 2$) with the experimental results at the $B$ factories~\cite{be:2004xa,be:2004data,ba:2005data}. The first error of our prediction is for $m_c\in [1.40\;{\rm GeV}, 1.60\;{\rm GeV}]$ and the second error is from the uncertainties of the wavefunction parameters $|R_{ns}(0)|$ and $|R^{'}_{np}(0)|$ determined by Eqs.(\ref{wfval1},\ref{wfval2},\ref{wfval3}). } \label{cross-m}
\end{table*}

\begin{table}[htb]
\begin{tabular}{|c|c|c|}
\hline
   &  $J/\psi +\chi_{c0}$  &  $\psi^{'} +\chi_{c0}$ \\
\hline Belle $\sigma \times{ B^{ \chi_{c0}}[> 2]}$ \cite{be:2004xa} & $16 \pm 5 \pm 4 $&  $17 \pm 8 \pm 7 $ \\
Belle $\sigma \times{ B^{ \chi_{c0}}[> 2(0)] }$ \cite{be:2004data} &
$6.4 \pm 1.7 \pm 1.0 $&  $12.5 \pm 3.8 \pm 3.1$ \\
BaBar $\sigma \times{ B^{ \chi_{c0}}[> 2] }$ \cite{ba:2005data}
& $10.3 \pm 2.5 ^{+1.4}_{-1.8} $& $\sim$ \\
\hline
Wang, Ma and Chao~\cite{W:2011nlo} & 9.5 & 4.1 \\
Dong, Feng and Jia~\cite{Dong:2011nlo} & 8.62 & 4.98 \\
\hline Our result & $12.25^{+3.70}_{-3.13}$ & $5.23^{+1.56}_{-1.32}$ \\
\hline
\end{tabular}
\caption{Total cross sections (in unit: fb) for $J/\psi (\psi^{'})+\chi_{c0}$ production at the $B$ factories. Our estimation together with those of Refs.\cite{W:2011nlo,Dong:2011nlo} are presented as a comparison. } \label{cross-n}
\end{table}

A comparison of our estimations with the experimental results are shown in Table \ref{cross-m} and \ref{cross-n}, where as a comparison, we also list the results of Refs.\cite{W:2011nlo,Dong:2011nlo}. The errors for our present estimations are listed in Table \ref{cross-m}, which are caused by varying $m_c\in [1.40\;{\rm GeV}, 1.60\;{\rm GeV}]$ and by considering the uncertainties of the wavefunction parameters $|R_{ns}(0)|$ and $|R^{'}_{np}(0)|$ determined by Eqs.(\ref{wfval1},\ref{wfval2},\ref{wfval3}). Our estimation for the total cross section of $J/\psi+\chi_{c0}$ production is consistent with the experimental result, but the $\psi'+\chi_{c0}$ production cross section is still smaller than the central value of the data. Since the data is still with large error, a future more accurate measurement shall be helpful to clarify this puzzle. As a final remark, there is residual scale dependence due to unknown higher order $\{\beta_i\}$-terms, which however is highly suppressed~\cite{pmc1,pmc2,pmc3,pmc4,pmc5}.

\begin{figure*}[tb]
\includegraphics[width=0.32\textwidth]{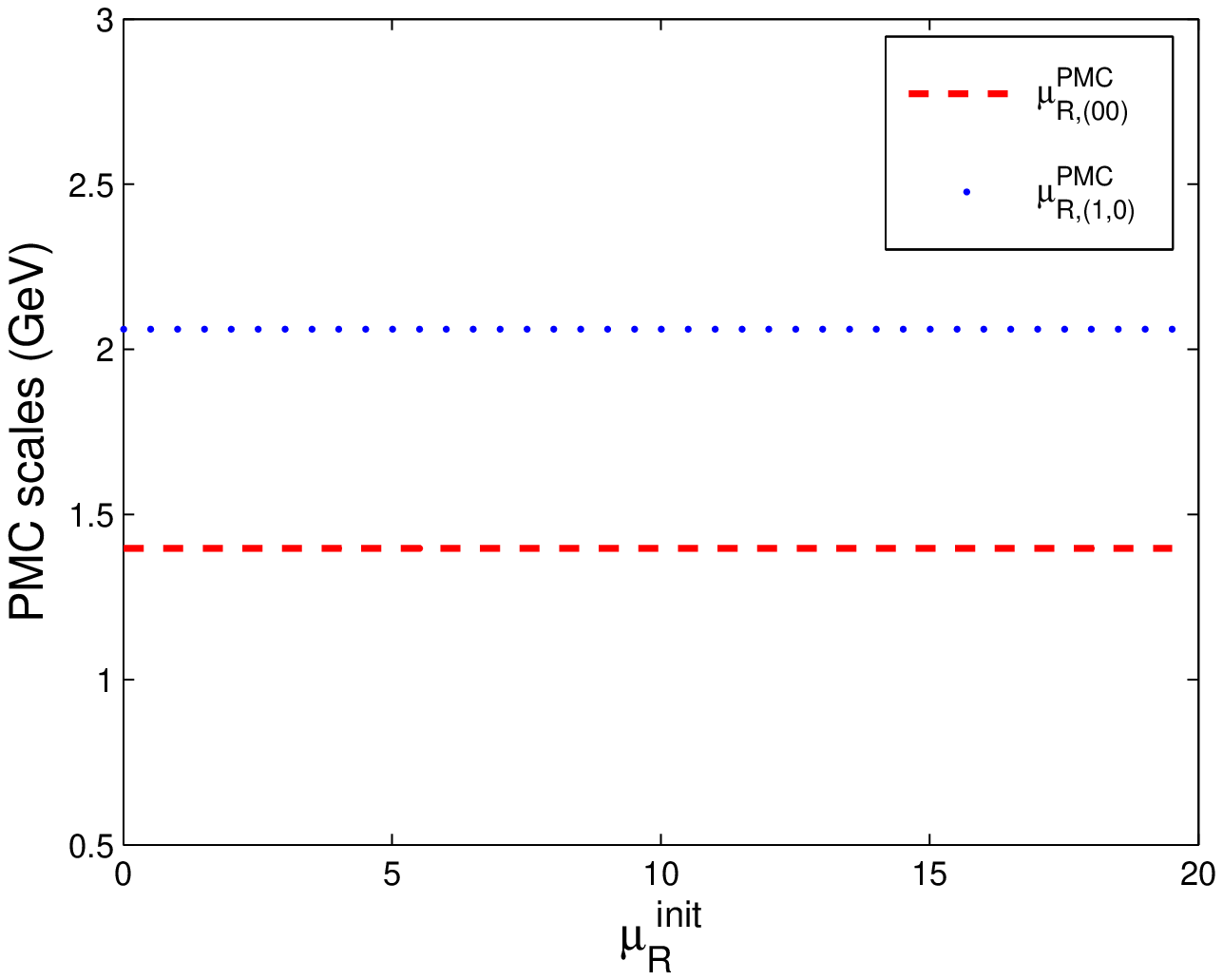}
\includegraphics[width=0.32\textwidth]{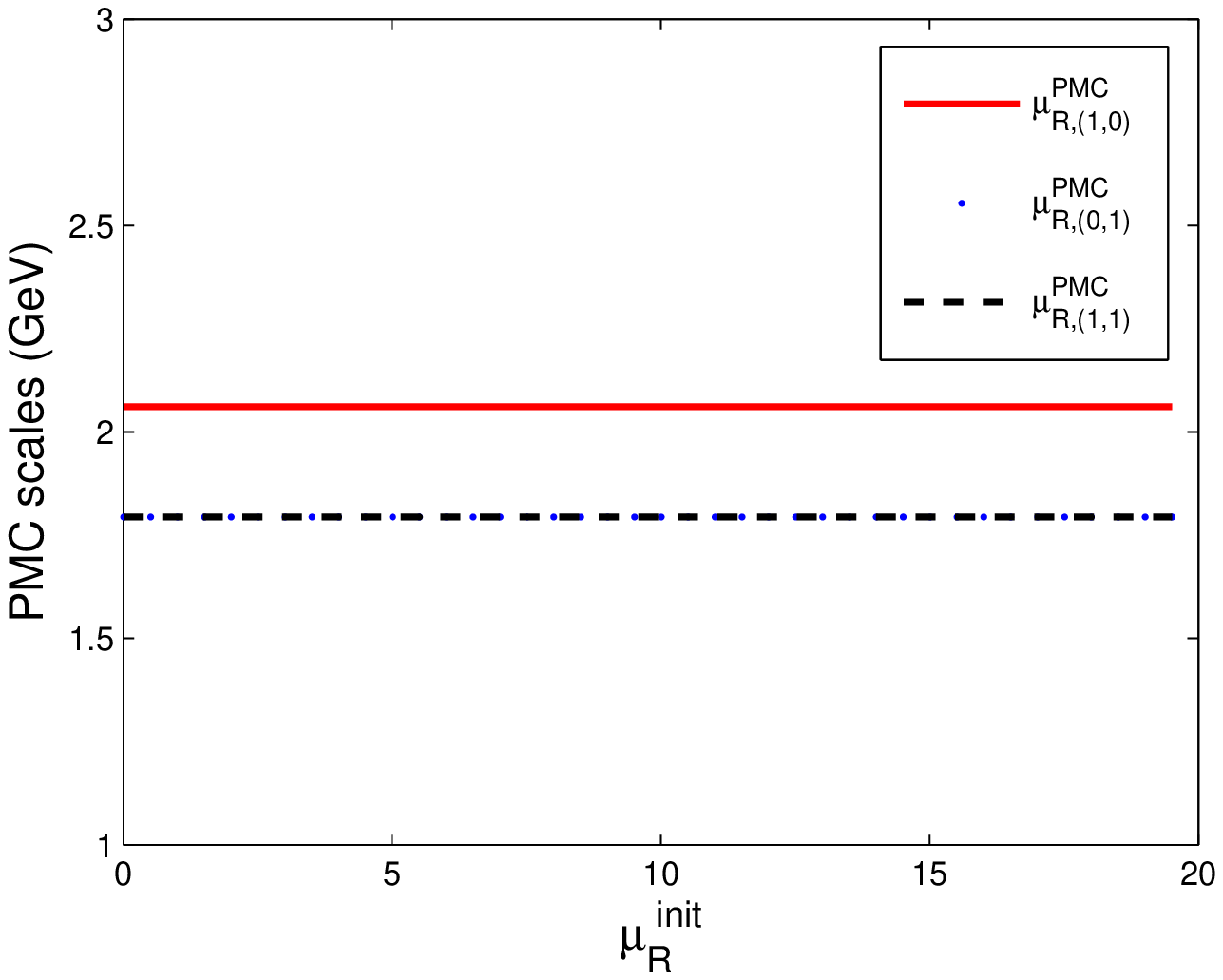}
\includegraphics[width=0.32\textwidth]{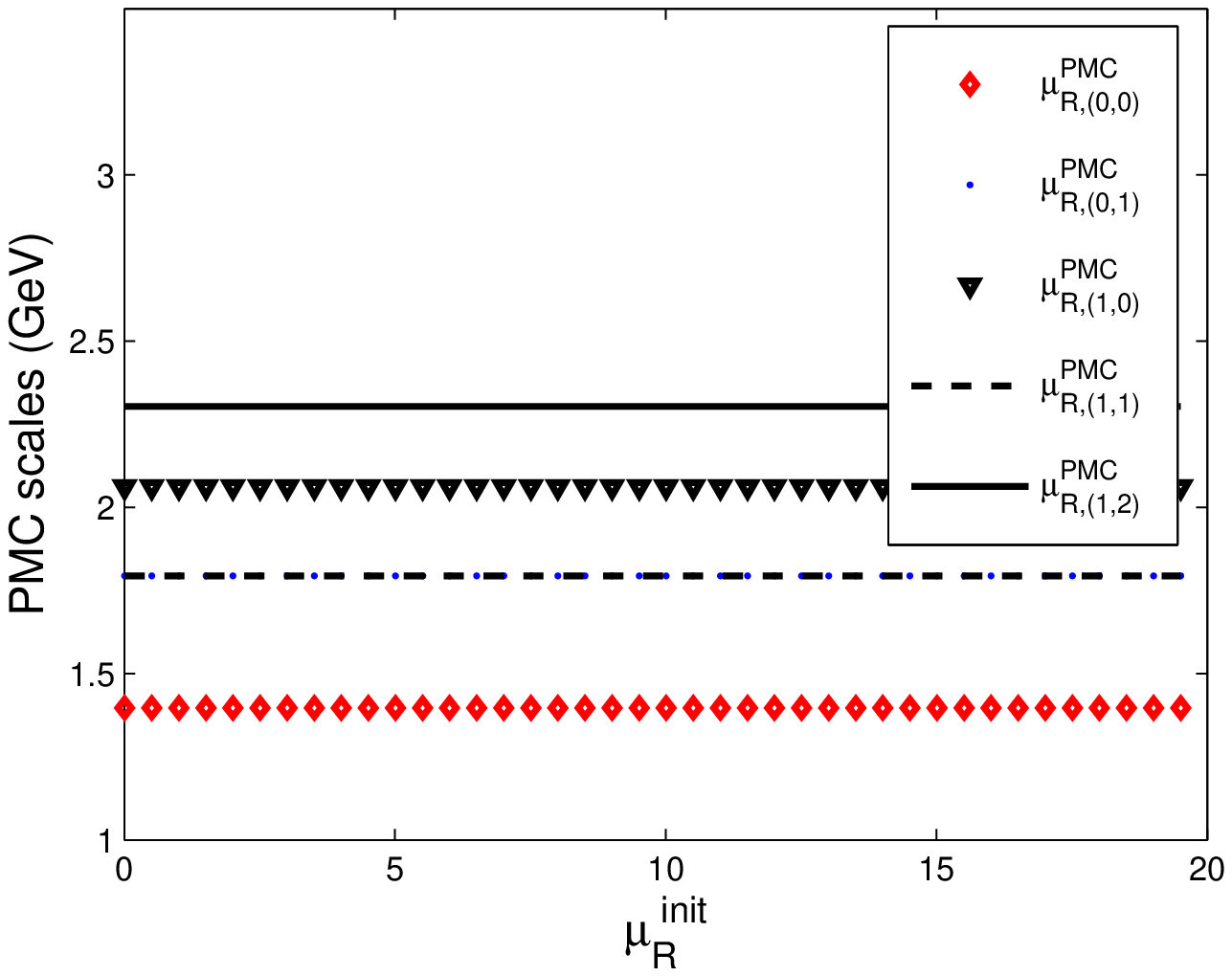}
\caption{PMC scales versus the initial renormalization scale $\mu^{\rm init}_{R}$ for the polarized cross sections of $e^{+}+e^{-}\rightarrow J/ \psi + \chi_{cJ}$ with $J=0,1,2$. } \label{Plot:mu:pmcscale}
\end{figure*}

In addition to the cases of the total cross sections, it is also interesting to show how PMC scale setting affects the polarized cross sections. We take $e^{+}+e^{-}\rightarrow J/ \psi + \chi_{cJ}$ as an explicit example. For each polarized process, the coefficients for the conformal or the non-conformal terms are different due to the enhancement or cancelation among different polarizations, then it is reasonable that the effective momentum flows of them are different. So, in principle, we should introduce different PMC scales for each polarized process.

More explicitly, we put the PMC scales for the polarized cross section versus the initial renormalization scale $\mu^{\rm init}_R$ in Fig.(\ref{Plot:mu:pmcscale}), where the PMC scales for $J=0,1,2$ are presented. All PMC scales for the polarized cross sections are almost independent of the initial renormalization scale, which are similar to the cases of total cross sections. Then, the scale uncertainty are also eliminated for the polarized cross sections at the NLO level. This also shows that our present calculation for both the polarized and unpolarized cross sections are consistent with each other. Fig.(\ref{Plot:mu:pmcscale}) shows
\begin{itemize}
\item For $e^{+}+e^{-}\rightarrow J/ \psi + \chi_{c0}$ process, we need to introduce two different PMC scales for $(\lambda_1,\lambda_2)=(0,0)$ and $(1,0)$, respectively. The contribution from the helicity channel $(1,0)$ is about $2$ times larger than that of the $(0,0)$ channel.
\item For $e^{+}+e^{-}\rightarrow J/ \psi + \chi_{c1}$ process, we need to introduce three PMC scales. The PMC scales for $(\lambda_1,\lambda_2)=(0,1)$ and $(1,1)$ are equal to each other due to the same non-conformal terms $B^{1(\beta)}_{0,1}=B^{1(\beta)}_{1,1}$, so we only have two different PMC scales. The helicity channel $(0,1)$ provides the dominant contribution for this process.
\item For $e^{+}+e^{-}\rightarrow J/ \psi + \chi_{c2}$ process, we need to introduce five PMC scales corresponding to five helicity channels. The PMC scales for $(\lambda_1,\lambda_2)=(0,1)$ and $(1,1)$ are the same due to the same non-conformal terms $B^{2(\beta)}_{0,1}=B^{2(\beta)}_{1,1}$, while other PMC scales are different, so we have four different PMC scales. The contribution from the helicity channels $(0,0)$ and $(1,0)$ are dominant over others.
\end{itemize}

\begin{table}[ht]
\begin{tabular}{|c||c|c|c|c|c|c|}
\hline
~ ~ & \multicolumn{3}{c|}{~Conventional scale setting~} & \multicolumn{3}{c|}{~PMC scale setting~} \\
\hline
~ ~ & ~LO~ & ~NLO~ & ~sum~ & ~LO~& ~NLO~& ~sum~ \\
\hline
~ $\sigma^{0}_{0,0}$ (fb)~ & 0.876 & 0.643 & 1.519 & 2.917 & 0.243 & 3.160 \\
\hline
$\sigma^{0}_{1,0}$ (fb) & 1.467 & 1.207 & 2.674 & 3.053 & 1.484 & 4.537\\
\hline
\end{tabular}
\caption{Polarized cross sections for $e^{+}+e^{-}\rightarrow J/ \psi + \chi_{c0}$ before and after PMC scale setting. $\mu^{\rm init}_R={\sqrt{s}/2}$. } \label{scaleun0}
\end{table}

\begin{table}[ht]
\begin{tabular}{|c||c|c|c|c|c|c|}
\hline
& \multicolumn{3}{c|}{Conventional scale setting} & \multicolumn{3}{c|}{PMC scale setting} \\
\hline
~ ~ & ~LO~ & ~NLO~ & ~sum~ & ~LO~& ~NLO~&~ sum \\
\hline
~$\sigma^{1}_{1,0}$ (fb)~ & 0.001 & -0.007& -0.006 & 0.002 & -0.023 & -0.021 \\
\hline
$\sigma^{1}_{0,1}$ (fb) & 0.294 & 0.114 & 0.408 & 0.709 & -0.187 & 0.522\\
\hline
$\sigma^{1}_{1,1}$ (fb) & 0.026 & -0.001 & 0.025 & 0.063 & -0.058 &0.005\\
\hline
\end{tabular}
\caption{Polarized cross sections for $e^{+}+e^{-}\rightarrow J/ \psi + \chi_{c1}$ before and after PMC scale setting. $\mu^{\rm init}_R={\sqrt{s}/2}$. } \label{scaleun1}
\end{table}

\begin{table}[ht]
\begin{tabular}{|c||c|c|c|c|c|c|}
\hline
& \multicolumn{3}{c|}{Conventional scale setting} & \multicolumn{3}{c|}{PMC scale setting} \\
\hline
~ ~ & ~LO~ & ~NLO~ & ~sum~ & ~LO~& ~NLO~&~ sum \\
\hline
~$\sigma^{2}_{0,0}$ (fb)~ & 0.343 & -0.050 & 0.293 & 1.142 & -1.739 & -0.597 \\
\hline
$\sigma^{2}_{1,0}$ (fb) & 0.212 & 0.191 & 0.403 & 0.442 & 0.262 & 0.704 \\
\hline
$\sigma^{2}_{0,1}$ (fb) & 0.051 & 0.003 & 0.054 & 0.123 & -0.094 & 0.029 \\
\hline
$\sigma^{2}_{1,1}$ (fb) & 0.026 & 0.004 & 0.030 & 0.063 & -0.041 &0.022 \\
\hline
$\sigma^{2}_{1,2}$ (fb) & 0.002 & 0.0003 & 0.002 & 0.003 & -0.001 &0.002 \\
\hline
\end{tabular}
\caption{Polarized cross sections for $e^{+}+e^{-}\rightarrow J/ \psi + \chi_{c2}$ before and after PMC scale setting. $\mu^{\rm init}_R={\sqrt{s}/2}$. } \label{scaleun2}
\end{table}

Because of the exponential suppression to the PMC scale as shown by Eq.(\ref{pmcscale}), the LO cross section for each polarized cross section shall be increased to a certain degree, which are shown clearly in Tables \ref{scaleun0}, \ref{scaleun1} and \ref{scaleun2}. Defining a parameter $C$, being the ratio of the cross section at NLO level to that at LO level, we find that the parameter $C$ are $0.73$ and $0.82$ for $(\lambda_1,\lambda_2)=(0,0)$ and $(1,0)$ channels of $e^{+}+e^{-}\rightarrow J/ \psi + \chi_{c0}$ under the conventional scale setting; which however changes down to $0.08$ and $0.49$ after PMC scale setting. For $e^{+}+e^{-}\rightarrow J/ \psi + \chi_{c1,2}$, the conditions are similar. This shows that the pQCD series become more convergent after PMC scale setting.

\section{summary}

We have applied PMC scale setting to study the polarized and the unpolarized cross sections of $e^{+}+e^{-}\rightarrow J/ \psi + \chi_{cJ}(J=0,1,2)$ up to NLO level. After PMC scale setting, we find that the final results are scale- and scheme- independent and the pQCD series becomes more convergent for both the polarized and the unpolarized cross sections. So we gain a more accurate pQCD estimation than those in the literature. More explicitly,
\begin{itemize}
\item After PMC scale setting, we obtain the following total cross sections
    \begin{eqnarray}
\sigma(J/\psi+\chi_{c0}) &=& 12.25^{+3.70}_{-3.13}\; {\rm fb},\;\nonumber\\
\sigma(J/\psi+\chi_{c1}) &=& 1.00^{+0.41}_{-0.32}\; {\rm fb},\;\nonumber\\
\sigma(J/\psi+\chi_{c2}) &=& 1.58^{+0.49}_{-0.43}\; {\rm fb}\nonumber
    \end{eqnarray}
    and
    \begin{eqnarray}
\sigma(\psi'+\chi_{c0})  &=& 5.23^{+1.56}_{-1.32}\; {\rm fb},\;\nonumber\\
\sigma(\psi'+\chi_{c1})  &=& 0.43^{+0.17}_{-0.14}\; {\rm fb},\;\nonumber\\
\sigma(\psi'+\chi_{c2})  &=& 0.67^{+0.22}_{-0.18}\; {\rm fb},\nonumber
   \end{eqnarray}
   where the errors are caused by taking the charm quark mass $m_c\in [1.40\;{\rm GeV}, 1.60\;{\rm GeV}]$ and the uncertainties of the wavefunction parameters $|R_{ns}(0)|$ and $|R^{'}_{np}(0)|$ determined by Eqs.(\ref{wfval1},\ref{wfval2},\ref{wfval3}). Our estimation for the total cross section of $J/\psi+\chi_{c0}$ production is consistent with the experimental result, but the $\psi'+\chi_{c0}$ production cross section is still far less than the central value of the $B$-factory data~\cite{be:2004xa,be:2004data,ba:2005data}. Since the data is still with large error, a future more accurate measurement shall be helpful to clarify this puzzle.

\item A comparison of Tables \ref{cross-m}, \ref{scaleun0}, \ref{scaleun1} and \ref{scaleun2} shows that $e^{+}+e^{-}\to J/ \psi (\psi') + \chi_{cJ}$ processes are dominated by $J=0$ channel, which have been measured at the $B$ factories. In contrast, the contribution of the $e^{+}+e^{-}\rightarrow J/ \psi (\psi') + \chi_{c1,c2}$ channels for this process seem to be rather modest even after PMC scale setting, which are about one order lower than that of $\chi_{c0}$ production case. The future super $B$ factory, with much higher luminosity, may eventually observe these polarization patterns.

\item Under conventional scale setting, the NLO contribution for all the channels are not small. After PMC scale setting, the pQCD convergence are greatly improved for most of the polarized and unpolarized channels. Thus the unknown NNLO corrections for those channels will not changes our present estimations much, since the contributions from those unknown NNLO $\beta$-terms shall be exponentially suppressed to the present PMC scales~\cite{pmc1,pmc2,pmc3}, which inversely guarantees the unknown conformal terms also provide less important contributions than that of the NLO conformal term. There are also cases when the pQCD convergence cannot be improved even after applying the PMC scale setting. For those cases, we really need a NNLO calculation to make the pQCD estimation more reliable, but which will also not heavily change our present estimations and conclusions. This is similar to the case of top-pair production at the hadronic colliders, where we have found that after the PMC scale setting, not only the scale dependence can be eliminated but also the pQCD convergence can be greatly improved only after finishing a NNLO calculation~\cite{pmc1,pmc3,pmc4}.

\item As a cross check, by taking all the same input parameters, we exactly obtain the results listed in Ref.\cite{Dong:2011nlo}. It is found that after PMC scale setting, even though we have introduced different PMC scales for different type of cross sections caused by the different non-conformal terms, by summing up all the polarized cross sections, we can get the same result as that of the directly calculated unpolarized cross section for the case of $J=0$ and $J=1$. This shows that our PMC procedure is self-consistent. However, there is slight difference for the case of $J=2$. This shows that the expression for the coefficient $B^{2(con)}_{0,0}$ listed in Ref.\cite{Dong:2011nlo} could have some typos for the conformal terms.

\item It is found that after PMC scale setting, for the polarized cross sections $\sigma^{1}_{1,0}$ (negative even before PMC scale setting) and $\sigma^{2}_{0,0}$, we obtain the ``incorrect" negative cross section. Similar to the QCD pomeron case, this could be the problem of the unphysical $\overline{\rm MS}$-scheme itself~\cite{pomeron1,pomeron2}. It has already been pointed that there is also the same ``incorrect" sign for the NLO highest eigenvalue of the BFKL equation under the $\overline{\mbox{MS}}$-scheme~\cite{pomq1,pomq2}. In Ref.\cite{pomeron1}, it has been shown that the reliability of QCD predictions for the intercept of the BFKL pomeron at NLO when evaluated using the physical schemes, such as the momentum space subtraction (MOM) scheme \cite{mom1,mom2,mom3}, can be significantly improved.

    The ``incorrect" sign for $\sigma^{1}_{1,0}$ and $\sigma^{2}_{0,0}$ under $\overline{\rm MS}$ scheme may show that those unknown non-conformal terms from the NNLO or higher orders should have sizable contributions to the LO PMC scale and can not be safely neglected. In fact, the large $K$ factor for these two polarized cross sections shows their pQCD convergence are terrible. It is found that when we transforming the cross sections from the $\overline{\rm MS}$ schemes to the MOM scheme, such ``incorrect" sign disappears. A detailed discussion on this point shall be presented elsewhere~\cite{prepare}.

\item As a final remark: $e^{+}+e^{-}\to J/ \psi (\psi') + \chi_{cJ}$ processes provide another good example for eliminating the renormalization scale ambiguity by using the PMC. Thus, after applying PMC scale setting, it will not only increase the precision of QCD tests, but also it will increase the sensitivity of the collider experiments to new physics beyond the standard model.

\end{itemize}

{\bf Acknowledgments :} The authors would like to thank F. Feng, W.L. Sang and Y. Jia for helpful discussions. This work was supported in part by Natural Science Foundation of China under Grant No.11075225 and No.11275280, by the Program for New Century Excellent Talents in University under Grant No.NCET-10-0882, and by Fundamental Research Funds for the Central Universities under Grant No.CDJXS12301102 and No.CQDXWL-2012-Z002. \\

\section*{Appendix: coefficients for the LO and NLO terms}

The expressions for the LO coefficients $c^J_{\lambda_1,\lambda_2}$ and the NLO coefficients $B^J_{\lambda_1,\lambda_2}(\mu_R)$ can be found in Ref.\cite{Dong:2011nlo}. Here, we correct some more typos in the non-conformal $\{\beta_0\}$-terms that are listed in Ref.\cite{Dong:2011nlo} by using explicit relations among the same polarization combination $(\lambda_1,\lambda_2)$ with different $J$ \footnote{Those relations are confirmed through private communications with the authors of Ref.\cite{Dong:2011nlo}.}. For convenience, we put all the LO coefficients $c^J_{\lambda_1,\lambda_2}(r)$ in the following
\begin{eqnarray}
c^0_{0,0}(r) &=& -1-10r+12r^2, \nonumber\\
c^0_{1,0}(r) &=& -9+14r, \nonumber\\
c^1_{0,1}(r) &=& \sqrt{6}(2-7 r), \nonumber\\
c^1_{1,0}(r) &=& -\sqrt{6}r, \nonumber\\
c^1_{1,1}(r) &=& 2\sqrt{6}(1-3r), \nonumber\\
c^2_{0,0}(r) &=& \sqrt{2}(-1+2r+12r^2), \nonumber\\
c^2_{0,1}(r) &=& \sqrt{6}(-1+5r), \nonumber\\
c^2_{1,0}(r) &=& \sqrt{2}(11r-3), \nonumber\\
c^2_{1,1}(r) &=& 2\sqrt{6}(1-3r), \nonumber\\
c^2_{1,2}(r) &=& -2\sqrt{3}. \nonumber
\end{eqnarray}

Using the asymptotic expressions derived in Ref.\cite{Dong:2011nlo} and the relations among the same polarization combination $(\lambda_1,\lambda_2)$ with different $J$, the non-conformal NLO coefficients $B^{J(\beta)}_{\lambda_1,\lambda_2}$ and the conformal NLO coefficients $B^{J(con)}_{\lambda_1,\lambda_2}$ at the renormalization scale $\mu_R$ can be written as:
\begin{eqnarray}
B^{0(\beta)}_{0,0} &=& B^{2(\beta)}_{0,0}={1 \over 2} \bigg(\ln\frac{\mu^2_R}{s} +{8\over 3}+ 2\ln2 \bigg), \\
B^{0(\beta)}_{1,0} &=& B^{1(\beta)}_{1,0}=B^{2(\beta)}_{1,0}={1 \over 2}\bigg( \ln\frac{\mu^2_R}{s} + {17\over 9} + 2\ln2 \bigg), \\
B^{1(\beta)}_{0,1} &=& B^{2(\beta)}_{0,1}={1 \over 12}\bigg(
6\ln\frac{\mu^2_R}{s} + 12\ln2 + 13 \bigg), \\
B^{1(\beta)}_{1,1} &=& B^{2(\beta)}_{1,1}= {1 \over 12}\bigg(6\ln\frac{\mu^2_R}{s} + 12\ln2 + 13 \bigg), \\
B^{2(\beta)}_{1,2} &=& {1 \over 2}\bigg( \ln\frac{\mu^2_R}{s} + {5\over 3} + 2\ln2 \bigg)
\end{eqnarray}
and
\begin{widetext}
\begin{eqnarray}
B^{0(con)}_{0,0} &=& -{2\over 3} (4-\ln2) \ln r - \frac{1}{9} ( 46 + \pi^2 - 40\ln2 + 33 \ln^2 2), \\
B^{0(con)}_{1,0} &=& {2\over 3} \ln^2 r - {1 \over 54}(139-104\ln2)\ln r - {1\over 27}\bigg( 161 + {8\pi^2\over 3}- {495\over 2}\ln2 + 100 \ln^2 2 \bigg), \\
B^{1(con)}_{1,0} &=& {-1\over 6 r} \bigg(5 \ln^2r + (7-2\ln2)\ln r-19+ 2\pi^2 + 75\ln2- 21 \ln^2 2\bigg), \\
B^{1(con)}_{0,1} &=& {1\over 12}\bigg( \frac{25}{2} \ln^2 r - (46-99\ln2)\ln r - {1\over 6} (616 + 74\pi^2 - 1696\ln2 + 303 \ln^22 )\bigg), \\
B^{1(con)}_{1,1} &=& {1\over 12}\bigg( 10 \ln^2r + 2(1+17\ln2)\ln r - \frac{1}{3}( 266 + 7\pi^2 - 128\ln 2 + 147 \ln^2 2)\bigg). \\
B^{2(con)}_{0,0} &=& -{2\over 3}(4-\ln2) \ln r -\frac{1}{9} ( 64 + \pi^2 + 104\ln 2 + 33 \ln^22), \\
B^{2(con)}_{0,1} &=& {1\over 6} \bigg( \frac{13}{2}\ln^2 r - (22-43\ln2) \ln r - \frac{1}{6} ( 284 + 30\pi^2 - 380\ln2 + 159 \ln^2 2)\bigg), \\
B^{2(con)}_{1,0} &=& -{1\over 3} \bigg(2\ln^2 r +\frac{1}{6}(5+8\ln2) \ln r  -\frac{1}{18} ( 291-8\pi^2 + 171 \ln 2 + 312\ln^2 2 )\bigg), \\
B^{2(con)}_{1,1} &=& {1\over 12} \bigg( 4\ln^2 r - (46-62\ln2) \ln r  - \frac{1}{3}(274 + 27\pi^2 - 316\ln 2 + 9\ln^2 2)\bigg), \\
B^{2(con)}_{1,2} &=& -{1\over 2} \bigg( 2\ln^2 r + \frac{2}{3}(1+13\ln2) \ln r + \frac{1}{9} ( -7\pi^2 + 140 - 104\ln 2 + 237\ln^2 2)\bigg).
\end{eqnarray}
\end{widetext}

\end{document}